\documentclass[11pt]{article}
\usepackage{amssymb}
\usepackage{color}
\title{\textbf{\textbf{ Group theoretical construction of Planar Noncommutative Phase Spaces }}}%Rotation-less Kinematical Symmetries in the planar
\author{Ancille Ngendakumana\footnote{nancille@yahoo.fr}\\Institut de
Math\'ematiques et des Sciences Physiques, Porto-Novo, Benin\\ and \\
Joachim Nzotungicimpaye\footnote{kimpaye@kie.ac.rw}
 \\Kigali Institute of Education, Kigali, Rwanda\\ and \\ Leonard Todjihound\'e\footnote{leonardt@imsp.uac.org}\\Institut de
Math\'ematiques et des Sciences Physiques,Porto-Novo,Benin\\}
\begin{document}
\maketitle
\date
%\tableofcontents
\begin{abstract}
Noncommutative phase spaces are generated and classified in the framework of centrally extended anisotropic planar kinematical Lie groups
as well as in the framework of noncentrally extended planar absolute time Lie groups.  Through these
constructions the coordinates of the phase spaces do not commute due to the presence of naturally introduced fields
giving rise to minimal couplings.
 By symplectic realizations methods, physical interpretations of generators coming from the obtained structures are given.\\
\end{abstract}

                              anisotropic kinematical groups, absolute time groups
\\
\\
\\
\\
\\
\\
\\
\tableofcontents
\section{Introduction}
Models associated with a given symmetry group can be conveniently constructed using Souriau's method. His theorem says
in fact that when a symmetry group $G$ acts transitively on a phase space, then the latter is a
coadjoint orbit of $G$ equipped with its canonical symplectic form \cite{souriau, kostant, kirillov}.
The first applications that Souriau presented in his book \cite{souriau} concern both the Poincar\'e and the Galilei groups for which coadjoint
orbits represent elementary particles characterized by the invariants $m$ (mass) and $s$ (spin).
Souriau himself goes one step further as he considers massless particles with spin, $m=0$, $s\neq0$ identified as relativistic and
nonrelativistic spin respectively. \\

 Souriau's ideas were later extended to larger groups. Taking \\$G=Poincare \times H_0$ where $H_0$ is an internal
symmetry group \\(e. g $SU(2), SU(3),...$) yields relativistic particles with internal structure (for more details see
\cite{duval2, duval-horvathy}).

The nonrelativistic kinematical groups admit nontrivial central extensions by one-dimensional algebra in dimension $d\geq 3$ but in the plane, they
admit an exotic \cite{horvathy6} two-parameter central extension.
The one-parameter central extension of the spatial Galilei group has been considered by Souriau in his book \cite{souriau}, the two-parameter
central extension of the planar Galilei group was studied in \cite{ horvathy6, horvathy7}.\\

Futhermore, Souriau's method has recently also been applied to smaller space-time symmetry groups .   For example, in \cite{duval3}
a classical \textquotedblleft photon \textquotedblright model was constructed,
 based entirely on the Euclidean group $E(3)$, a subgroup of both the Poincar\'e and the Galilei groups.
An other application of the Souriau's method is found in \cite{gonera} where the most general dynamical systems on which the
nonrelativistic conformal groups act transitively as symmetries are constructed. \\

Equivalently to the Souriau's theorem, the dual ${\cal{G}}^*$ of the Lie algebra $\cal{G}$ of $G$ has a natural
Poisson structure whose symplectic leaves are the coadjoint orbits.  Depending on the Lie group,
these orbits may provide noncommutative phase spaces.
Physical theories with noncommuting coordinates have become
the focus of recent research (see, e.g., \cite{ horvathy6, horvathy7,
grigore, horvathy2},...), the notion of noncommutativity having different physical interpretations.
For example, it is well known that velocities do not commute in the presence of an electromagnetic field.  Also, it has been proved
that in the presence of the dual electromagnetic field the positions do not commute  and
the maximal coadjoint orbits of the noncentrally extended planar Galilei and planar Para-Galilei groups have been shown to be models of
noncommutative phase spaces \cite{ancilla1}.\\

Futhermore, a phase space with noncommutative positions and noncommutative momenta has been realized by
Souriau's method on the anisotropic Newton-Hooke groups \cite{ancilla}. In \cite{zhang-horvathy}, the authors have found a similar
symmetry in the so-called Hill
problem, which is effectively an anisotropical harmonic oscillator in a magnetic field.  This system has no rotational symmetry while
translations and generalized boosts still act as symmetries.
%It also has a two-parameter central extension.
The noncommutative version of the Hill problem was discussed in \cite{zhang-horvathy1}.\\

%  In \cite{ancille}, the authors have used Souriau's method to construct
%  phase spaces equipped with modified symplectic structure on Aristotle group by using both the noncentral extension and its corresponding central
% extension.

In this paper, we use the Souriau's approach to construct and classify phase spaces of planar noncommutative systems.  We consider the case where
the symmetry groups are the kinematical groups \cite{5bacry} and realize noncommutative phase spaces on their maximal coadjoint orbits by using
central extensions of all anisotropic kinematical Lie algebras (when rotation invariance is released) .
Note that for the one-parameter centrally extended kinematical Lie algebras, the
 nontrivial Lie bracket which contains the only central extension parameter $m$ is
\begin{eqnarray}
 [K_i,P_j]=m\delta_{ij},
\end{eqnarray}
 which means that the generators of space translations as well as pure kinematical group transformations
commute.  One can not then
associate noncommutative phase spaces to the one-parameter centrally extended kinematical Lie groups.
  It
is then the absence of the symmetry rotations (i.e anisotropy of the plane)
 which
%implies a two-parameter centrally extended kinematical Lie group and
 guaranties the noncommutative phase
space for the anisotropic kinematical groups.  However it is possible to associate a noncommutative phase space to absolute time groups by
 considering
 their noncentral extensions.  Thus, we enlarge this theory by considering also
the noncentral extensions of the absolute time kinematical Lie algebras associated to the Lie groups classified in
\cite{mcrae}.  \\

Explicitly, we show that noncommutative symplectic structures can be generated in
the framework of centrally extended anisotropic kinematical Lie algebras as well as in the framework of
noncentrally extended isotropic kinematical Lie algebras (rotations included).
 However,
noncommutative phase spaces realized with
noncentral extensions of the kinematical Lie groups are geometrically more general than those constructed on their central extensions.
Furthermore, it is also shown in this paper that the noncommutativity of momenta implies
some modification of the second Newton law \cite{ancilla1, romero, wei}.
As the coadjoint orbit construction has not been curried through some of these planar kinematical Lie groups before,
physical interpretations of new generators of those extended structures are given.
In all these cases, the noncommutativity is measured by naturally introduced fields, each
 corresponding to a minimal coupling.\\
% and nonrelativistic classical
% elementary systems which share the rotation-less Newton-Hooke symmetry (\cite{zhang-horvathy},\cite{ancille}) are pointed out. \\

 The paper is organized as follows.
In section two, we give a review of the eleven possible kinematical Lie algebras.  We then extract those for which rotation generators can be
dropped producing anisotropic kinematical Lie algebras \cite{3derome}.  In the third section, we compute both central and noncentral
extensions of
the planar anisotropic kinematical Lie algebras and their corresponding Lie groups with the assumption that an abelian extension (central
or noncentral) of a Lie algebra should integrates to an abelian extension (central or noncentral) of its corresponding Lie group \cite{hekmati}.
In the fourth and fifth sections, we respectively summarize the coadjoint orbit method and
 construct, for the first time, the coadjoint orbits of both the centrally extended planar anisotropic and the
noncentrally extended absolute time kinematical Lie algebras.  We finally classify the noncommutative phase spaces obtained.

\section{Possible anisotropic kinematical groups}
Consider a manifold $M$ on which a transformation group $G$ acts transitively (this action is usually the left one) meaning that
$M$ is a $G$-homogeneous space.  When $M$ describes space-time, $G$ is the kinematical group of $M$.  All $d$-dimensional space-times
with a constant curvature have a kinematical group of dimension $\frac{1}{2}d(d+1)$ \cite{4defkingroup}.
In \cite{5bacry}, Bacry and L\'evy-Leblond have classified the possible ten-parameter kinematical groups consisting of
the space-time translations, spatial rotations and inertial transformations connecting different inertial frames of reference.
 Bacry and L\'evy-Leblond have shown, under the assumption that
\begin{itemize}
 \item the space must be isotropic, meaning that the rotation group $SO(d-1)$ generated by $J_a~,a=1,2,...,\frac{(d-2)(d-1)}{2}$, is a subgroup of
the kinematical group,
 \item the space-time must be homogeneous, meaning that the space translations group generated by $P_i,i=1,2,...,d-1$ and the time translations
 group generated by $H$, are subgroups of the kinematical group,
\item the inertial transformations group generated by $K_i,~i=1,2,...,d-1$ is a noncompact subgroup of the kinematical group,
\item the parity ($\pi:H\rightarrow H,P_i\rightarrow -P_i,K_i\rightarrow-K_i,J_a\rightarrow J_a$) and the time-reversal
($\theta:H \rightarrow -H,P_i\rightarrow P_i,K_i\rightarrow-K_i,J_a\rightarrow J_a$) are automorphisms of the kinematical group,
\end{itemize}
that there are eleven kinematical groups.
 If $L$ and $T$ denote respectively the dimension of a length and of a duration, then the physical dimensions of $P_i$, $H$ and $K_i$
 are $L^{-1}$, $T^{-1}$ and $L^{-1}T$ respectively. It is also known that the generators $J_a$ of rotations are dimensionless.\\

Their corresponding kinematical Lie algebras
are characterized by the fact that the inertial transformation generators and
the space translation generators behave as vectors under rotations while the time translation generator behaves as a scalar :
\begin{eqnarray}\label{rotations}
[J_j,J_k]=J_l\epsilon_{jk}^{l},~[J_j,K_k]=K_l\epsilon_{jk}^{l},~[J_j,P_k]=P_l\epsilon_{jk}^{l},~[J_j,H]=0
\end{eqnarray}
We also have that:
\begin{eqnarray}\label{boosts}
[K_j,K_k]=\mu J_l\epsilon_{jk}^{l},~[K_j,P_k] =\gamma\delta_{jk}H,~[K_j,H]=\lambda P_j
\end{eqnarray}
where the physical dimension of the parameters $\mu$ and $\gamma$ is $L^{-2}T^2$ while the parameter $\lambda$ is dimensionless. Finally,
\begin{eqnarray}\label{translations}
 [P_j,P_k] =\alpha J_l\epsilon_{jk}^l,~[P_j,H]=\beta K_j
\end{eqnarray}
where the physical dimension of the parameter $\alpha$ is $L^{-2}$ while that of the parameter $\beta$ is $T^{-2}$.\\
Only three of the five parameters $\alpha$, $\beta$, $\lambda$, $\mu$, $\gamma$ are independent.  Effectively the Jacobi identities
\begin{eqnarray*}
[K_i,[P_j,P_k]]+[P_j,[P_k,K_i]]+[P_k,[K_i,P_j]]=0
\end{eqnarray*}
and
\begin{eqnarray*}
[K_i,[K_j,P_k]]+[K_j,[P_k,K_i]]+[P_k,[K_i,K_j]]=0~
\end{eqnarray*}
imply that
\begin{eqnarray*}
\alpha=\beta\gamma~,~\mu=-\lambda\gamma
\end{eqnarray*}
If we compute the adjoint representation of the generators $K_i$, we verify that the non compacity of the boost
transformations imply that $\mu \leq 0$, meaning that $\lambda$ and $\gamma$ are all positive or all negative when one of
them is not equal to zero.  The brackets (\ref{boosts}) and (\ref{translations}) become
\begin{eqnarray*}\label{boosts1}
[K_j,K_k]=-\lambda\gamma J_l\epsilon_{jk}^{l},~[K_j,P_k] =\gamma\delta_{jk}H,~[K_j,H]=\lambda P_j
\end{eqnarray*}
and
\begin{eqnarray}\label{translations1}
 [P_j,P_k] =\beta\gamma J_l\epsilon_{jk}^l,~[P_j,H]=\beta K_j
\end{eqnarray}
We remain with three parameters $\beta$, $\gamma$ and $\lambda$ constrained by the fact that $\lambda$ and $\gamma$
are of the same sign when they are all different from zero.  If each generator is multiplied by $-1$, the three parameters change sign. As
$\lambda$ is dimensionless, we can assume ( after normalization ) that $\lambda=1$ or $\lambda=0$ and then that $\gamma \geq 0$. \\

Let $\kappa$
 denotes the space curvature and let $\omega$ denotes the time curvature (frequency).  Then, for each of the two values of $\lambda$,
$\gamma=\frac{1}{c^2}$ or
$\gamma=0$ where $c=\frac{\omega}{\kappa}$ is a velocity.  Also for each of the two values of $\gamma$, there are three
Lie algebras corresponding to
$\beta=\pm \omega^2$ and $\beta=0$.
\\We then
distinguish the case where boosts do not commute with time translations ( i.e $[K_i,H]=P_i)$ from the case where they commute ( i.e $[K_i,H]=0 $)
as detailed below.
\subsection{Boosts not commuting with time translations }
This case corresponds to $\lambda=1$. The kinematical Lie algebras are then defined by the brackets (\ref{rotations}), (\ref{translations1}) and
\begin{eqnarray*}\label{boosts2}
[K_j,K_k]=-\gamma J_l\epsilon_{jk}^{l},~[K_j,P_k] =\gamma\delta_{jk}H,~[K_j,H]= P_j
\end{eqnarray*}
According to the values of $\gamma$ and $\beta$, the possible planar kinematical Lie algebras of this form are summarized in the following
table where $d{\cal{S}}_+, ~d{\cal{S}}_-,~ {\cal{P}},$\\${\cal{NH}}_+,~{\cal{NH}}_-$ and $\cal{G}$ stand respectively for the de Sitter, the
 anti de Sitter, the
Poincar\'e, the expanding Newton-Hooke, the oscillating  Newton-Hooke and the Galilei Lie algebras.
\\
\\
\begin{table}[htbp]
{\bf \caption{ \it Kinematical Lie algebras whose boosts do not commute with time translations ( $[K_i,H]=P_i)$.}}
 \begin{tabular}{|c|c|c|}
\hline
$\gamma=\frac{1}{c^2}$&$~$&$[K_j,K_k]=-\frac{1}{c^2} J_l\epsilon_{jk}^{l},~[K_j,P_k] =\frac{1}{c^2}\delta_{jk}H,~[K_j,H]= P_j$\\
\hline
$d\cal{S}_+$&$\beta=\omega^2$&$[P_j,P_k] =\kappa^2 J_l\epsilon_{jk}^l,~~[P_j,H]=\omega^2 K_j$\\
\hline
$\cal{P}~~~~$&$\beta=0$&$[P_j,P_k] =0,~~~~~~~~~~~[P_j,H]=0~~~~~$\\
\hline
$d\cal{S}_-$&$\beta=-\omega^2$&$~~~[P_j,P_k] =-\kappa^2 J_l\epsilon_{jk}^l,~[P_j,H]=-\omega^2 K_j$\\
\hline
$\gamma=0$&$~$&$[K_j,K_k]=0$,$[K_j,P_k] =0$,$[K_j,H]= P_j$\\
\hline
$\cal{NH}_+$&$\beta=\omega^2$&$~~~[P_j,P_k] =0,~~~~~~~~~~~~~[P_j,H]=\omega^2 K_j$\\
\hline
$\cal{G}~~~$&$\beta=0$&$~~[P_j,P_k] =0,~~~~~~~~~~~~~~[P_j,H]=0$\\
\hline
$\cal{NH}_-$&$\beta=-\omega^2$&$~~~~~~~~~[P_j,P_k] =0,~~~~~~~~~~~[P_j,H]=-\omega^2 K_j$\\
\hline
 \end{tabular}

\end{table}

\subsection{Boosts commuting with time translations }
This case corresponds to $\lambda=0$.  The kinematical Lie algebras are then defined by the brackets (\ref{rotations}), (\ref{translations1}) and
\begin{eqnarray*}\label{boosts3}
[K_j,K_k]=0,~[K_j,P_k] =\gamma\delta_{jk}H,~[K_j,H]=0
\end{eqnarray*}
In this case, we obtain the kinematical Lie algebras summarized in the table below where ${\cal{P}}^{\prime}_+,~{\cal{P}}^{\prime}_-,~
{\cal{C}},~
{\cal{G}}^{\prime}_+,~{\cal{G}}^{\prime}_-$ and ${\cal{S}}$ stand respectively for the Para-Poincar\'e, the anti Para-Poincar\'e, the Carroll, the
Para-Galilei, the
Anti-Para-Galilei and the Static Lie algebras.
\begin{table}[htbp]
{\bf \caption{\it Kinematical Lie algebras whose boosts and time translations commute ( $[K_i,H]=0)$. }}
 \begin{tabular}{|c|c|c|}
\hline
$\gamma=\frac{1}{c^2}$&$~$&$[K_j,K_k]=0,~[K_j,P_k] =\frac{1}{c^2}\delta_{jk}H,~[K_j,H]= 0$\\
\hline
$\cal{P}^{\prime}_+$&$\beta=\omega^2$&$[P_j,P_k] =\kappa^2 J_l\epsilon_{jk}^l,~~[P_j,H]=\omega^2 K_j$\\
\hline
$\cal{C}~~$&$\beta=0$&$[P_j,P_k] =0,~~~~~~~~~~~[P_j,H]=0~~~~~$\\
\hline
${P}^{\prime}_- $&$\beta=-\omega^2$&$~~~[P_j,P_k] =-\kappa^2 J_l\epsilon_{jk}^l,~[P_j,H]=-\omega^2 K_j$\\
\hline
$\gamma=0$&$~$&$[K_j,K_k]=0~,~[K_j,P_k] =0,~~[K_j,H]= 0$\\
\hline
$\cal{G}^{\prime}_+$&$\beta=\omega^2$&$~~~[P_j,P_k] =0,~~~~~~~~~~~~~[P_j,H]=\omega^2 K_j$\\
\hline
$\cal{S}~~$&$\beta=0$&$~~[P_j,P_k] =0,~~~~~~~~~~~~~~[P_j,H]=0$\\
\hline
$\cal{G}^{\prime}_-$&$\beta=-\omega^2$&$~~~~~~~~~[P_j,P_k] =0,~~~~~~~~~~~~~[P_j,H]=-\omega^2 K_j$\\
\hline
 \end{tabular}
\end{table}
\\
In conclusion, the Lie brackets for the possible kinematical Lie algebras according to the classification in \cite {5bacry} are summarized
in the following table where the brackets
of the form $[J,X_i]=X_j\epsilon^j_i$ and $[J,H]=0$ are omitted.
\\
\begin{table}[htbp]
{\bf \caption{\it Lie brackets for the possible kinematical Lie algebras according to the classification in \cite{5bacry}. }}
\begin{tabular}{|c|c|c|c|c|c|c|}
\hline
Lie algebra& $[K_i,H]$&$[K_i,K_j]$&$[K_i,P_j]$&$[P_i,P_j]$& $[P_i,H]$\\
\hline
 $d\cal{S}_+$ & $P_i$ & $-\frac{1}{c^2}J_k\epsilon^k_{ij}$ & $\frac{1}{c^2}H\delta_{ij}$ & $\kappa^2J_k\epsilon^k_{ij}$& $\omega^2 K_i$ \\
\hline
  $\cal{P}$ & $P_i$ & $-\frac{1}{c^2}J_k\epsilon^k_{ij}$ & $\frac{1}{c^2}H\delta_{ij}$ & $0$& $0$ \\
\hline
 $d\cal{S}_-$ & $P_i$ & $-\frac{1}{c^2}J_k\epsilon^k_{ij}$ & $\frac{1}{c^2}H\delta_{ij}$ & $-\kappa^2J_k\epsilon^k_{ij}$&
$-\omega^2 K_i$ \\
\hline
  $\cal{NH}_+$ & $P_i$ & $0$ & $0$ & $0$& $\omega^2 K_i$ \\
\hline
 $\cal{G}$& $P_i$ & $0$ & $0$ & $0$& $0$ \\
\hline
$\cal{NH}_-$ & $P_i$ & $0$ & $0$ & $0$& $-\omega^2 K_i$ \\
\hline
$\cal{P}^{\prime}_+$&  $0$ & $0$ & $\frac{1}{c^2}H\delta_{ij}$ & $\kappa^2J_k\epsilon^k_{ij}$& $\omega^2 K_i$ \\
\hline
  $\cal{C}$& $0$ & $0$ & $\frac{1}{c^2}H\delta_{ij}$ & $0$& $0$ \\
\hline
 $\cal{P}^{\prime}_-$ & $0$ & $0$ & $\frac{1}{c^2}H\delta_{ij}$ & $-\kappa^2J_k\epsilon^k_{ij}$& $-\omega^2 K_i$ \\
\hline
  $\cal{G}^{\prime}$&  $0$ & $0$ & $0$ & $0$& $\omega^2 K_i$ \\
\hline
 $\cal{S}$ & $0$ & $0$ & $0$ & $0$& $0$ \\
 \hline
\end{tabular}
\end{table}
\\
\\
Futhermore, their corresponding Lie groups are distributed according to the table below:
\begin{table}[htbp]
{\bf\caption{ \it Kinematical groups classification according \cite{mcrae}}}
\begin{tabular}{|c|c|}
\hline
Relative time groups&de Sitter, Poincar\'e, Para-Poincar\'e, Carroll\\
\hline
Absolute time groups &Newton-Hooke, Galilei, Para-Galilei, Static\\
\hline
Relative space groups&de Sitter, Newton-Hooke, Poincar\'e, Galilei\\
\hline
Absolute space groups&Para-Poincare, Para-Galilei, Carroll, Static\\
\hline
Cosmological groups&de Sitter, Newton-Hooke, Para-Poincar\'e, Para-Galilei\\
\hline
Local groups&Poincar\'e, Galilei, Carroll, Static\\
\hline
\end{tabular}
\end{table}
\\
The corresponding anisotropic Lie algebras are obtained by dropping the generators $J_i$ of rotations, meaning that
they are generated by $\{K_i,P_i,H\}$. This is only possible for the two \textbf{Newton-Hooke} Lie algebras, the \textbf{Galilei} and
 the \textbf{Para-Galilei} Lie algebras, the \textbf{Carroll} Lie algebra and the \textbf{Static} Lie algebra where the rotation
generators do not appear in the right hand side of the brackets $[K_i,K_j]$ and $[P_i,P_j]$.  Except the Carroll group, all possible
anisotropic groups are absolute time groups.
Thus, next to these anisotropic kinematical Lie algebras we consider also the absolute time Lie algebras
 with respect to the isotropy of the two-dimensional space
because both the two types of kinematical Lie
algebras admit extensions.  In the following section, we determine central and noncentral extensions of the above planar kinematical Lie algebras.
 Hereafter, absolute (relative) time groups will be called nonrelativistic (relativistic) kinematical groups.

\section{Extensions of the planar kinematical Lie algebras}
\subsection{Central extensions of anisotropic kinematical Lie algebras}
\subsubsection{Relativistic anisotropic Lie algebra (Carroll Lie algebra)}
The Carroll Lie algebra, was first introduced in \cite{levy} as time-velocity contraction
\cite{inonu} of the Poincar\'e algebra through a rescaling of the boosts and time translations.  Although appearing naturally in the
 classification of kinematical groups, as an
alternative intermediate algebra in the contraction of the Poincar\'e group onto the Static group, and therefore as another limit (the other
being the Galilei algebra), the Carroll algebra has played no distinguish role in Kinematics.  However, recently it has been analyzed whether
this algebra constitutes an object in the study of the problem of tachyon condensation in string theory \cite{21}.  It is in this context
 where a possible
 cosmological interpretation of this limit of the Poincar\'e algebra and the noncommutative phase spaces on the Carroll group recover some interest.
\\

Furthermore, Carroll Lie algebra is the only relativistic algebra which is anisotropic.  Its only nontrivial Lie bracket is given by :
\begin{eqnarray}\label{carrolbracket}
[K_i,P_j]=\frac{1}{c^2}H\delta_{ij}
\end{eqnarray}
By using standard methods \cite{ kostant, kirillov, 5bacry, hamermesh,
 nzo1}, we obtain that the central extension of this Lie algebra is defined by the following Lie brackets
\\
\begin{eqnarray}\label{carrolbrackets}
 [K_i,K_j]=\frac{1}{c^2}S\epsilon_{ij}~~~~~~~~~~~~~~~~~~~~~~~~~~~~~~~~~~~~\\\nonumber
[K_i,P_j]=\frac{1}{c^2}H\delta_{ij},~[P_i,P_j]=\kappa^2 S\epsilon_{ij}~~~~~~~~~~~~\\\nonumber
[K_i,H]=0,~~~~[P_i,H]=0~~~~~~~~~~~~~~~~~~~~~~~~\nonumber
\end{eqnarray}
where $S$ generates the center of this extended Lie algebra and is dimensionless while $\kappa$ is the space curvature.
\subsubsection{Absolute time anisotropic Lie algebras}
The cohomological structure which determines the existence of central \\extensions of absolute time kinematical groups originates in
their invariant subgroup spanned by translations and boosts \cite{souriau}.\\
The five nonrelativistic anisotropic Lie algebras are defined by the following Lie structure :
\begin{eqnarray*}\label{anisotropicnr}
[K_j,K_k]=0~~~~~~~~~~~~~~~~~~~~~~~~~~\\ \nonumber
[K_j,P_k] =0,~[P_j,P_k] =0~~~~~~~~~\\\nonumber
[K_j,H]=\lambda P_j,~[P_j,H]=\beta K_j ~~\nonumber
\end{eqnarray*}
with $\lambda=1$ or zero, $\beta=\pm \omega^2$ or zero.\\Dimensional analysis permit us to set that a priori
the possible central \\extensions are defined by the following Lie brackets :
\\
\begin{eqnarray}\label{anisotropicext}
[K_j,K_k]=\frac{\mu}{c^2}S\epsilon_{jk}~~~~~~~~~~~~~~~~~~~~~~~~~~~~~~\\ \nonumber
[K_j,P_k] =\gamma M\delta_{jk},~[P_j,P_k] =\kappa^2\alpha S\epsilon_{jk}~~~~~~\\\nonumber
[K_j,H]=\lambda P_j,~[P_j,H]=\beta K_j~~~~~~~~~~~~~~~~ \nonumber
\end{eqnarray}
where $S$ , $\alpha$, $\gamma$ and $\mu$ are dimensionless while the dimension of $M$ is $L^{-2}T$
and that of the parameter $\beta$ is $L^{-2}$.\\ The Lie brackets (\ref{anisotropicext})
will form Lie algebras if every triplet satisfies the Jacobi identity.  The Jacobi identities
\begin{eqnarray*}
 [K_i,[P_j,H]]+[P_j,[H,K_i]]+[H,[K_i,P_j]]=0
\end{eqnarray*}
 imply that
\begin{eqnarray}\label{relation}
 \frac{\mu\beta}{c^2}=-\kappa^2 \lambda\alpha
\end{eqnarray}
By using relation $c=\frac{\omega}{\kappa}$ in (\ref{relation}) for the above possible values of $\lambda$ and $\beta$, we get the following cases :
\begin{itemize}
 \item $\mu=\mp\alpha$ for $\lambda=1$ and $\beta=\pm\omega^2$
\item $\mu\in \Re$ while $\alpha =0$ when $\lambda=1$ and $\beta=0$
\item $\alpha\in\Re$ while $\mu=0$ when $\lambda=0$ and $\beta=\omega^2$
\item$\mu,\alpha \in \Re$ when $\lambda=0$ and $\beta=0$
\end{itemize}
 We can assume after normalization ($\mu=1$ and $\gamma=1$)
that the central extended planar nonrelativistic anisotropic Lie algebras are given by the Lie brackets summarized in the following table:
\\

\begin{table}[htbp]
{\bf\caption{\it Central extensions of nonrelativistic anisotropic kinematical Lie algebras}}
\begin{tabular}{|c|c|c|c|c|c|c|}
\hline
Lie algebra&$[P_i,H]$&$[K_i,H]$&$[P_i,P_j]$&$[K_i,K_j]$& $[K_i,P_j]$\\
\hline
 Extended $NH_+$ & $\omega^2 K_{i}$& $P_i$ & $-\kappa^2 S\epsilon_{ij}$ & $\frac{1}{c^2}S\epsilon_{ij}$&$M\delta_{ij}$ \\
\hline
 Extended $NH_-$& $-\omega^2 K_i$ & $P_i$ & $\kappa^2 S\epsilon_{ij}$ &$\frac{1}{c^2}S\epsilon_{ij}$ & $M\delta_{ij}$ \\
\hline
 Extended $G~~~~$ & $0$ & $P_i$ & $0$ & $\frac{1}{c^2}S\epsilon_{ij}$ & $M\delta_{ij}$ \\
\hline
Extended $G^{\prime}_{\pm}~$ & $\pm\omega^2 K_{i}$ & $0$ & $\kappa^2 S\epsilon_{ij}$  & $0$ & $M\delta_{ij}$ \\
\hline
Extended $S~~~~ $& $0$ & $0$ &  $\kappa^2 S\epsilon_{ij}$ & $\frac{1}{c^2}S\epsilon_{ij}$ & $M\delta_{ij}$ \\
\hline
\end{tabular}
\end{table}
These centrally extended anisotropic Lie algebras are new except for the Newton-Hooke groups case \cite{ancilla, zhang-horvathy, galajinski}.
\subsection{Noncentral extensions of the absolute time planar kinematical Lie algebras}
The six absolute time Lie algebras are defined by the following
nontrivial Lie brackets
\begin{eqnarray*}
 [J,K_j]=K_i\epsilon^i_j,~[J,P_j]=P_i\epsilon^i_j,~[K_i,H]=\lambda P_i,~[P_i,H]=\beta K_i,~i,j=1,2
\end{eqnarray*}
with $\lambda=1$ or zero, $\beta=\pm \omega^2$ or zero.  We distinguish four cases :
 \subsubsection{Newton-Hooke Lie algebras}
The Newton-Hooke Lie algebras correspond to the case $\lambda=1$ and \\$\beta=\pm \omega^2$. Dimensional analysis permits us to set that
after normalization
the noncentral extensions of $\cal{NH}_{\pm}$ coincides with their two-fold centrally extended Lie algebras whose nontrivial Lie brackets are
 given by:
\begin{eqnarray*}\label{anisotropicextnc}
[J,K_j]=K_i\epsilon^i_j,~[J,P_j]=P_i\epsilon^i_j,~[K_j,K_k]=\frac{1}{c^2}S\epsilon_{jk},~[K_j,P_k] = M\delta_{jk}
\end{eqnarray*}
and
\begin{eqnarray*}\label{anisotropicextnc1}
[P_j,P_k] =\kappa^2 S\epsilon_{jk}~,[K_j,H]= P_j,~[P_j,H]=\pm \omega^2 K_j~~~~~~~~~~~~~~~~
\end{eqnarray*}
where $S$ and $M$ generate the center of these centrally extended algebras, ${\kappa}$ is a constant space curvature
while $\omega$ is the time curvature.\\
\subsubsection{Para-Galilei Lie algebras}
The Para-Galilei Lie algebras correspond to the case $\lambda=0$ and $\beta=\pm \omega^2$.
Their corresponding noncentrally extended Lie algebras has the following structure
 \begin{eqnarray*}\label{anisotropicextg}
[J,K_j]=K_i\epsilon^i_j,~[K_j,K_k]=0~~~~~~~~~~~~~~~~~~~~~~~~~~~~~~~~~~~~~\\\nonumber
[J,P_j]=P_i\epsilon^i_j,~[K_j,P_k] = M\delta_{jk},~[P_j,P_k] =\kappa^2 S\epsilon_{jk}~~~~~~~~\\\nonumber
 [J,H]=0,~~[K_j,H]= \Pi_j~~,~[P_j,H]=\pm \omega^2 K_j~~~~~~~~~~~~~ \\\nonumber
[J,\Pi_j]=\Pi_i\epsilon^i_j~~~~~~~~~~~~~~~~~~~~~~~~~~~~~~~~~~~~~~~~~~~~~~~~~~~~~~~\\\nonumber
\end{eqnarray*}
where $\vec{\Pi}$ behaves as a vector under rotations and has the dimension as that of $\vec{P}$ and where $S$ and $M$ commute with other generators
of this noncentrally extended Lie algebras.
This case corresponds to the planar noncentrally extended Para-Galilei Lie algebras, cfr \cite{ancilla1}.
\subsubsection{Galilei Lie algebra}
The Galilei Lie algebra corresponds to the case $\lambda=1$ and $\beta=0$.  Its noncentrally extended Lie algebra has the following Lie structure :
\begin{eqnarray*}
[J,K_j]=K_i\epsilon^i_j~,~~[K_j,K_k]=\frac{1}{c^2}S\epsilon_{jk}~~~~~~~~~~~~~~~~~~~~~~~~~~~~~~\\\nonumber
[J,P_j]=P_i\epsilon^i_j~,~~~[K_j,P_k] = M\delta_{jk},~[P_j,P_k] =0~~~~~~~~~~~~~~~\\\nonumber
[J,F_j]=F_i\epsilon^i_j, [K_i,F_j]=0,[P_i,F_j]=0~~~~~~~~~~~~~~~~~~~~~~~~~~~\\\nonumber
[J,H]=0,~[K_j,H]= Pi_j,~~[P_j,H]=F_j~~~~~~~~~~~~~~~~~~~~~~~\\\nonumber
\end{eqnarray*}
This noncentrally extended Lie structure has been used in \cite{ancilla1}.
\subsubsection{Static Lie algebra}
The Static Lie algebra corresponds to the case $\lambda=0$ and $\beta=0$.
Its noncentral extension has the following Lie structure :
 \begin{eqnarray}\label{anisotropicextg}
[J,K_j]=K_i\epsilon^i_j~,~~[K_j,K_k]=0~~~~~~~~~~~~~~~~~~~~~~~~~~~~~~\\\nonumber
[J,P_j]=P_i\epsilon^i_j~,~~~[K_j,P_k] = M\delta_{jk},~[P_j,P_k] =0~~~~~~~\\\nonumber
[J,F_j]=F_i\epsilon^i_j,~[K_j,F_k]=B\delta_{jk},~[P_j,F_k]=\Lambda\delta_{jk}~~~~~~\\\nonumber
[J,\Pi_j]=\Pi_i\epsilon^i_j,~~[K_j,\Pi_k] = M^{\prime}\delta_{jk},~~ [P_j,\Pi_k] =B\delta_{jk}~\\\nonumber
[J,H]=0,~[K_j,H]= \Pi_j,~~[P_j,H]=F_j~~~~~~~~~~~~~~~~\\\nonumber
\end{eqnarray}
This result is quite new and corresponds to the planar noncentrally extended Static Lie algebra.

\section{Coadjoint orbits}
The notion of {\it coadjoint orbits} is the main ingredient of the orbit method.  Moreover it is the most important mathematical object that has
been brought into consideration in connection with the orbit method.  Indeed,
this method relates the coadjoint orbit of a group $G$ to a phase space.
The construction of the coadjoint orbits and their canonical symplectic structures (Kirillov-Kostant-Souriau) is presented in detail
own to describe examples of noncommutative phase spaces on kinematical Lie groups. \\

Let $G$ be a Lie group and ${\cal{G}}$ its Lie algebra.  Let $Ad:G\rightarrow Aut({\cal{G}})$ be the adjoint representation
of $G$ on its Lie algebra ${\cal{G}}$ such that the automorphism $Ad_g$ associated to $g \in G$ is defined by
\begin{eqnarray*}
 Ad_g(X)=g Xg^{-1},~ X\in {\cal{G}}
\end{eqnarray*}
If ${\cal{G}^*}$ is the dual of ${\cal{G}}$, it is well known that the coadjoint action of $G$ on ${\cal{G}}^*$
$Ad^*:{\cal{G}}^*\times {\cal{G}} \rightarrow \Re$ is such that:
\begin{eqnarray*}
  \langle Ad^*_X(\alpha),Y)\rangle=\langle\alpha,[X,Y]\rangle
\end{eqnarray*}
If $\alpha=\alpha_i\epsilon^i \in {\cal{G}}^*,~~X=e_iX^i,~~Y=e_iY^i \in \cal{G}~~$
then
\begin{eqnarray*}
  \langle Ad^*_X(\alpha),Y)\rangle=K_{ij}(\alpha)X^iY^j
\end{eqnarray*}
where
\begin{eqnarray}\label{kirillov}
 K_{ij}(\alpha)=\alpha_kC_{ij}^k
\end{eqnarray}
is the Kirillov $2$-form \cite{levyleblond} on $\cal{G}^*.$
The representation $\rho:G\rightarrow {\cal{F}}({\cal{G}^*})$ of ${\cal{G}}$ on the space of vector fields on ${\cal{G}}^*$ defined by
 \begin{eqnarray*}
 \rho(X_i)=K_{ij}(\alpha)\frac{\partial }{\partial \alpha_j}
\end{eqnarray*}
is a Lie algebra homomorphism such that
$$ Ker(K(\alpha))=\{f \in C^{\infty}({\cal{G}}^*,\Re): \rho(X)f=0, X\in {\cal{G}}\}.$$
This means that $ Ker(K(\alpha))$ is the set of all invariants $f$ of $ {\cal{G}}$ in ${\cal{G}}^*$ satisfying the following relation:
\begin{eqnarray}\label{kirillovsystem}
  K_{ij}(\alpha)\frac{\partial f}{\partial \alpha_j}=0
 \end{eqnarray}
The quotient space $${\cal{O}}_{\alpha}^*=
{\cal{G}}^*/Ker(K(\alpha)),$$ called the coadjoint orbit of $G$ in ${\cal{G}}^*$, is a symplectic manifold \cite{giachetti} whose symplectic
form $\sigma^{ij}$ is obtained from
$$ \Omega_{ij}\sigma^{jk}=\delta^k_{i}$$
where $\Omega_{ij}=K_{ij}\backslash {\cal{O}}_{\alpha}^*$, i.e the restriction of the Kirillov form to the orbit.
Explicitly, the two-symplectic form is given by the following relation
 \begin{eqnarray}\label{symplectictwo-form}
 \sigma=(\Omega^{-1})^{ab}dx_b\wedge dx_a
 \end{eqnarray}
which takes the form $\sigma=dp_i\wedge dq^i$ in the canonical coordinates.\\

%  In conclusion, a coadjoint orbit is a symplectic manifold whose dimension corresponds to the rank of the Kirillov form while the symplectic form
% is defined by the
%  inverse of the restriction of the Kirillov form on the orbit.\\

If $x^a=(p_i,q^i)$, the Poisson bracket implied by the Kirillov symplectic structure
\begin{eqnarray}\label{poissonbrackets}
\{H,f\}=-\Omega_{ab}\frac{\partial H}{\partial
x_a}\frac{\partial f}{\partial x_b}
\end{eqnarray}
leads to
\begin{eqnarray}\label{canonicalcoordinates}
\{p_k,p_i\}=0~~,~~\{p_k,q^i\}=\delta^i_k~,~\{q^k,q^i\}=0
\end{eqnarray}
where $ p_i$, $q^i$ represent the generalized canonical coordinates and momenta of the system.  Relations (\ref{canonicalcoordinates}) mean
that the momenta commute within themselves as well as the
positions.\\

Interesting consequences arise by considering central and noncentral extensions of Lie algebras and this provides a more general symplectic
 two-form (\ref{symplectictwo-form}) whose extended Poisson brackets are given by
\begin{eqnarray*}\label{poissonext}
 \{x_{a},x_{b}\}=\Theta_{ab}
\end{eqnarray*}
 where
\begin{eqnarray*}
\Theta=\left(\begin{array}{cccc}
0&G&1&0\\-G&0&0&1\\-1&0&0&F\\0&-1&-F&0
\end{array}
\right)
\end{eqnarray*}
is the inverse of the matrix of the symplectic form
\begin{eqnarray*}
\Omega=\frac{1}{1-GF}\left(\begin{array}{cccc}
0&F&-1&0\\-F&0&0&-1\\1&0&0&G\\0&1&-G&0
\end{array}
\right)
\end{eqnarray*}
The fields $F$ and $G$ are constant because they are coming from central or noncentral extensions of Lie algebras.  But cases where they are not
constant have been considered \cite{acatrinei}.
Moreover the respective
physical dimensions of $G^{ij}$ and $F_{ij}$ are $M^{-1}T$ and $MT^{-1}$, $M$ representing a mass while $T$ represents a time.\\

The noncommutative phase space is then
 defined as a space on which variables satisfy the commutation relations:
 \begin{eqnarray*}
 \{ q^i,q^j \}=G^{ij}~~,~~   \{ p_i,q^j \}=\delta^{j}_i~~,~~ \{ p_i,p_j \}= F_{ij}
\end{eqnarray*}
where $\delta^{j}_i$ is a unit matrix.\\
The equations of motion corresponding to the above symplectic structure are given by:
\begin{eqnarray*}
 \dot{x^i}=\{H,x^i\}=\Theta^{ij}\frac{\partial H}{\partial x^j}
\end{eqnarray*}
more explicitly:
\begin{eqnarray}\nonumber\label{motionequations}
 \dot{q^i}=\frac{\partial H}{\partial p_i}+G\epsilon^{ij}\frac{\partial H}{\partial x^j}\\
\dot{p_i}=-\frac{\partial H}{\partial q^i}+F\epsilon_{ij}\frac{\partial H}{\partial p_j}
\end{eqnarray}
If $G=F=0$, then (\ref{motionequations}) are the usual Hamiltonian equations.\\

\section{ Noncommutative phase spaces constructed group \\theoretically}
Planar noncommutative phase spaces are constructed on both anisotropic kinematical groups by working with their central extensions
(subsection $5.1$)
and on absolute time groups with respect to the isotropy of the two-dimensional space by considering their noncentral extensions
(subsection $5.2$).  As it has been said in the introduction,
 the authors in \cite{ancilla1} and \cite{ancilla} have constructed noncommutative phase spaces by coadjoint orbit method starting with
the noncentrally extended Galilei and Para-Galilei Lie algebras and the
centrally extended anisotropic Newton-Hooke Lie algebras respectively.  In
this section, we construct noncommutative phase spaces on the other planar kinematical Lie algebras.
\subsection{ Noncommutative phase spaces on anisotropic kinematical groups}
\subsubsection{Newton-Hooke noncommutative phase space}
  Nonrelativistic particle models
have been constructed following Souriau's method for the two-parameter centrally extended anisotropic Newton-Hooke groups in a
two-dimensional space \cite{ancilla}
yielding similar results as in \cite{zhang-horvathy}. Indeed, by considering the oscillating Newton-Hooke group, the authors in
\cite{zhang-horvathy} have found a similar symmetry in the so-called Hill problem (the latter is studied in celestial mechanics), which is
effectivelly
an anisotropic harmonic oscillator in a magnetic field.  The peculiarity is that this system has no rotational symmetry while translations and
boosts still act as symmetries.  Note also that, as already said in the introduction, the noncommutative
version of the Hill problem was been discussed in \cite{zhang-horvathy1}. For all these reasons
this case is not reviewed in this section.

\subsubsection{Galilean noncommutative phase space}
The ($2+1$)-Galilei group $G$ is a six-parameter Lie group.  It is the kinematical group of a classical, nonrelativistic space-time having two
 spatial and one time dimensions.  It consists of translations of time and space, rotations and velocity boosts.
The planar Galilei group $G$ and its Lie algebra ${\cal{G}}$ have been explicitly defined in \cite{ancilla1}.
The latter can also be obtained from the Poincar\'e algebra
through the velocity-space contraction defined by rescaling the boosts and the space translation generators \cite{inonu}. \\

The central extension of the corresponding anisotropic Galilei Lie algebra is defined by the brackets given in the table ($5$)
where $M$ and $S$ generate its center, $c$
being a constant velocity.\\

Let
$k_iK^{*i}+p_iP^{*i}+EH^*+mM^*+hS^*$ be the general element of
the dual of the planar centrally extended  Lie algebra
where $\vec{k}$ is a kinematic momentum,
$\vec{p}$ is a linear momentum, $E$ is an energy, $m$ is a mass and
$h$ is an action.  Then $m$ and $h$ are trivial invariants under the coadjoint action of the planar anisotropic Galilei group.  The
other invariant, the solution of the Kirillov's system (\ref{kirillovsystem}), is explicitly given by:
\begin{eqnarray}\label{galileaninvariants}
U=e-\frac{\vec{p}~^2}{2m}
\end{eqnarray}
and is interpreted as the internal energy \cite{souriau}.
The inverse $\Omega^{-1}$ of the restriction $\Omega$ of the Kirillov's matrix on the orbit is given by
% \begin{eqnarray*}
% \Omega=\left(
% \begin{array}{cccc}
% 0&\frac{h}{c^2}&m&0\\
% -\frac{h}{c^2}&0&0&m\\
% -m&0&0&0\\
% 0&-m&0&0\\
% \end{array}
%  \right)
% \end{eqnarray*}
\begin{eqnarray*}
\Omega^{-1}=\left(
\begin{array}{cccc}
0&0&-\frac{1}{m}&0\\0&0&0&-\frac{1}{m}\\\frac{1}{m}&0&0&\frac{1}{m\omega_0}\\0&\frac{1}{m}&-\frac{1}{m\omega_0}&0
\end{array}
\right)
\end{eqnarray*}
where $\omega_0$ is defined by
\begin{eqnarray}\label{planck}
h\omega_0=mc^2
\end{eqnarray}
a relation remembering us the wave-particle duality, the left hand
side being an energy associated to a frequency, the right hand side
being an energy associated to a mass.\\

By using the relation
\begin{eqnarray}\label{kqm}
 q^i=\frac{k^i}{m},
\end{eqnarray}
and the wave-particle duality (\ref{planck}), we obtain that
 the Poisson bracket (\ref{poissonbrackets}) takes the form:
\begin{eqnarray*}\label{galileipoisson}
\{H,f\}=\frac{\partial H}{\partial p_i}\frac{\partial f}{\partial
q^i}-\frac{\partial H}{\partial q^i}\frac{\partial f}{\partial
p_i}+G^{ij}\frac{\partial H}{\partial q^i}\frac{\partial f}{\partial
q^j}
\end{eqnarray*}
with
\begin{eqnarray*}\label{g}
G^{ij}=-\frac{\epsilon^{ij}}{m\omega_0}
\end{eqnarray*}

Futhermore, the maximal coadjoint orbit denoted by
${\cal{O}}_{(m,h,U)}$
is equipped with the modified symplectic $2$-form:
\begin{eqnarray} \sigma=dp_i\wedge
dq^i+\frac{\epsilon^{ij}}{m\omega_0}dp_i\wedge dp_j
\end{eqnarray}

 which corresponds to the minimal coupling of position with the naturally introduced dual magnetic potential \cite{ancilla}
 where coordinates are:
\begin{eqnarray*}
\pi_i=p_i~~,~~x^i=q^i+\frac{p_k}{2m\omega_0}\epsilon^{ki}
\end{eqnarray*}
It follows that
\begin{eqnarray*}
\{p_i,p_j\}=0~~,~~\{p_i,x^k\}=\delta^k_i~~,~~ \{x^i,x^j\}=G^{ij}
\end{eqnarray*}

So following the coadjoint orbit method, we have constructed a nonrelativistic particle model for
 the two-parameter centrally extended anisotropic Galilei group in a two-dimensional space, recovering the exotic model described in
 \cite{horvathy7}.  It is a noncommutative phase space whose positions do not commute. This noncommutativity is due to presence of a naturally
 introduced dual magnetic field given by the relation
\begin{eqnarray*}
e^*B^*=-\frac{1}{m\omega_0}
\end{eqnarray*}
Moreover
the equations of motion corresponding to the above symplectic structure are given by:
\begin{eqnarray*}\label{garahamilton}
\frac{dp_i}{dt}=-\frac{\partial H}{\partial
q^i}~~,~~\frac{dx^i}{dt}=\frac{\partial H}{\partial
p_i}+\frac{\epsilon^{ki}}{2m\omega_0}\frac{\partial H}{\partial q^k}
\end{eqnarray*}
i.e.
\begin{eqnarray*}
\frac{dp_i}{dt}=-\frac{\partial H}{\partial
q^i}~~,~~\frac{dx^i}{dt}=\frac{\partial H}{\partial
p_i}-\frac{\epsilon^{ki}}{2m\omega_0}\frac{dp_k}{dt}
\end{eqnarray*}
 It has been proved in \cite {walczyk} that for Hamiltonian function of the form :
\begin{eqnarray}\label{hamiltonian0}
 H=\frac{1}{2m}(p_1^{2}+p_2^{2})+V(x^1,x^2),~~V(x^1,x^2)=\sum_iF_ix^i,~~F_i= \mbox{const.}
\end{eqnarray}
the corresponding Newton equation
\begin{eqnarray}\label{newtonlaw}
m \frac{d^2 x_i}{dt^2}=F_i
\end{eqnarray}
remains undeformed.\\
In the following section, we realize the Poisson brackets of the form :
\begin{eqnarray*}\label{un}
\{p_k,p_i\}=F_{ki}~,~\{p_k,q^i\}=\delta^i_k~,~\{q^k,q^i\}=0
\end{eqnarray*}
by the coadjoint orbit method on the planar anisotropic Para-Galilei group and prove that the noncommutativity of momenta implies
the modification of the second Newton law (\ref{newtonlaw}) \cite{romero, wei}.\\

\subsubsection{Para-Galilean noncommutative phase space}
The nonrelativistic Para-Galilei group is obtained through a space-time contraction of the Newton-Hooke groups or through a space-velocity
contraction of the Para-Poincar\'e group.  It contracts itself by a velocity-time contraction in the Static group \cite{5bacry, inonu}.\\

The planar Para-Galilei group and its Lie algebra have been explicitly defined in \cite{ancilla1}.
The central extension of its corresponding anisotropic Lie algebra generated by ${K_i,P_i,H}$ is defined by the Lie brackets given in the
table ($5$) where $M$ and $S$ generate the center of its planar centrally extended Lie algebra.\\

Let $k_iK^{*i}+p_iP^{*i}+EH^*+mM^*+hS^*$ be a general element of
the dual of the centrally extended anisotropic Para-Galilei Lie algebra where
$\vec{k}$ is a kinematic momentum,
$\vec{p}$ is a linear momentum, $E$ is an energy, $m$ is a mass and
$h$ is an action.\\

% Then the Kirillov form (\ref{kirillov}), is in this case given by :
% \begin{eqnarray*}
% (K_{ij})=\left(
% \begin{array}{ccccc}
% 0&0&m&0&0\\
% 0&0&0&m&0\\
% -m&0&0&\kappa^2h&\omega^2k_1\\
% 0&-m&-\kappa^2h&0&\omega^2k_2\\
% 0&0&-\omega^2k_1&-\omega^2k_2&0\\
% \end{array}
%  \right)
% \end{eqnarray*}
The coadjoint orbit of the Para-Galilei centrally extended Lie group on the dual of its Lie algebra
is then characterized by the two trivial invariants $m$ and $h$, and a
nontrivial invariant $U$, solution of the system (\ref{kirillovsystem}) and given by :
\begin{eqnarray}\label{para-galileaninvariants}
U=E+\frac{m\omega^2\vec{q}~^2}{2}
\end{eqnarray}
where we have used relation $c=\frac{\omega}{\kappa}$, (\ref{planck}) and (\ref{kqm}).  Let us denote by
${\cal{O}}_{(m,h,U)}$ the coadjoint orbit.\\The inverse $\Omega^{-1}$ of
the restriction $\Omega=(\Omega_{ab})$ of the Kirillov form (in the basis $(K_1,K_2,P_1,P_2,H)$) on ${\cal{O}}_{(m,h,U)}$ is then
\begin{eqnarray*}
\Omega^{-1}=\left(
\begin{array}{cccc}
0&\frac{\omega^2}{m\omega_0}&-\frac{1}{m}&0\\-\frac{\omega^2}{m\omega_0}&0&0&-\frac{1}{m}\\\frac{1}{m}&0&0&0\\0&\frac{1}{m}&0&0
\end{array}
\right)
\end{eqnarray*}
where we have used the wave-particle duality and the relation $c=\frac{\omega}{\kappa}$.\\
% \begin{eqnarray*}
% \Omega=\left(
% \begin{array}{cccc}
% 0&0&m&0\\
% 0&0&0&m\\
% -m&0&0&\kappa^2h\\
% 0&-m&-\kappa^2h&0\\
% \end{array}
% \right)
% \end{eqnarray*}
The Poisson bracket (\ref{poissonbrackets}) is in this case given by:
\begin{eqnarray*}\label{para-galileipoisson}
\{H,f\}=\frac{\partial H}{\partial p_i}\frac{\partial f}{\partial
q^i}-\frac{\partial H}{\partial q^i}\frac{\partial f}{\partial
p_i}+F_{ij}\frac{\partial H}{\partial p_i}\frac{\partial f}{\partial
p_j}
\end{eqnarray*}
with
\begin{eqnarray*}
F_{ij}=-\frac{\epsilon^{ij} m\omega^2}{\omega_0}
\end{eqnarray*}
 where we have used the relation the wave-particle duality (\ref{planck}) and (\ref{kqm}).\\
Moreover, the symplectic form (\ref{symplectictwo-form}) takes the form :
\begin{eqnarray}\label{para-galsigma}
 \sigma=dp_i\wedge
dq^i+\frac{\epsilon^{ij}m\omega^2}{\omega_0}dq^i\wedge dq^j
\end{eqnarray}
If the frequency of the charged particle still unchanged : i.e $\omega_0=\omega$, this symplectic structure is equivalent to the one obtained in
\cite{zhang-horvathy}.
\\
 The (modified) symplectic structure (\ref{para-galsigma}) corresponds to the minimal coupling of momenta with magnetic potential \cite{ancilla}
where the coordinates
\begin{eqnarray*}\label{magneticcoupling1}
 \pi_i=p_i+\frac{\epsilon^{ij} m \omega^2}{2 \omega_0} q^k~~,~~q^i=x^i
\end{eqnarray*}
satisfy
\begin{eqnarray*}
\{x^i,x^k\}=0~~~,~~~\{\pi_i,x^k\}=\delta^k_i~~,~~~\{\pi_i,\pi_k\}= F_{ik}
\end{eqnarray*}\\
So with the planar anisotropic Para-Galilei group, we have
obtained noncommutative phase space whose momenta do not commute.  This noncommutativity is due to the presence of a naturally introduced
magnetic field $B$ given by the relation:
\begin{eqnarray}\label{magnetic}
e B\epsilon^{ij}=F_{ij}
\end{eqnarray}
The corresponding equations of motion take the form:
\begin{eqnarray*}
\frac{dp_i}{dt}=-\frac{\partial H}{\partial
q^i}+F_{ij}\frac{\partial H}{\partial p_i}\delta^i_{j}~~,~~\frac{dx^i}{dt}=\frac{\partial H}{\partial
p_i}
\end{eqnarray*}
i.e.
\begin{eqnarray*}
\frac{dp_i}{dt}=-\frac{\partial V}{\partial
q^i}-\frac{\epsilon^{ij} m \omega^2}{\omega_0}\frac{dx^j}{dt}~~~,~~\frac{dx^i}{dt}=\frac{\partial H}{\partial
p_i}
\end{eqnarray*}
or equivalently
\begin{eqnarray}\label{secondnewtonlaw}
 m\frac{d^2 x^i}{dt^2}=-\frac{\partial V}{\partial{x^i}}+eB \epsilon^{ij}\frac{p_j}{m}
\end{eqnarray}
for Hamiltonian of the form (\ref{hamiltonian0}) and where we have used (\ref{magnetic}).
We interpret the equation (\ref{secondnewtonlaw}) as the modified Newton's second law \cite{romero, wei}.  The second term in this equation is
a correction due to
the noncommutativity of momenta. It is a damping force which depends on the space through the factor of noncommutativity $F_{ij}$.  For $F_{ij}=0$
or $B=0$, equation (\ref{secondnewtonlaw}) leads to the usual Newton's second law.

\subsubsection{Anisotropic Static noncommutative phase spaces }
The planar Static Lie algebra is an abelian Lie algebra generated by $J$ for rotation,
 $\vec{K}$ for boosts, $\vec{P}$ for space translations and $H$ for
time translations.  The general element of the connected Static group can be written as $$g=\exp(\vec{v}\vec{K}+\vec{x}\vec{P}+t H)\exp(\theta J)$$
where the parameter $\vec{v}$, $\vec{x}$ and $t$ are respectively the velocity parameter, the space translations parameter and the time translations
parameter, $\theta$ is an angle.
We restrict our study to the planar anisotropic Static group.\\

Let consider the central extension of the planar anisotropic Static Lie algebra whose Lie algebra stucture is given in the
table ($5$).\\

Let
$hS^*+k_iK^{*i}+p_iP^{*i}+EH^*+mM^*$ be the general element of
the dual of the planar anisotropic centrally extended Static Lie algebra. Then $E$, $m$ and $h$ are
trivial invariants under the coadjoint action of the planar anisotropic Static Lie group. \\
The restriction of the Kirillov's matrix on the orbit is given by
\begin{eqnarray*}
\Omega=\left(
\begin{array}{cccc}
0&\frac{h}{c^2}&m&0\\-\frac{h}{c^2}&0&0&m\\-m&0&0&\kappa^2h\\0&-m&-\kappa^2h&0
\end{array}
\right)
\end{eqnarray*}
By using the wave-particle duality (\ref{planck}) and the equality $c=\frac{\omega}{\kappa}$,
we obtain that the Poisson bracket of two functions implied by the Kirillov symplectic structure is given by
\begin{eqnarray*}
\{h,f\}=\frac{\partial h}{\partial p_i}\frac{\partial f}{\partial
q^i}-\frac{\partial h}{\partial q^i}\frac{\partial f}{\partial
p_i}+G^{ij}\frac{\partial h}{\partial q^i}\frac{\partial f}{\partial
q^j}+F_{ij}\frac{\partial h}{\partial p_i}\frac{\partial f}{\partial
p_j}~~;~~i,j=1,2
\end{eqnarray*}
with
\begin{eqnarray*}
G^{ij}=-\frac{\epsilon^{ij}}{m\omega_0}~~,~~ F_{ij}=-(m-\mu_{e})\omega\epsilon_{ij}
\end{eqnarray*}
and where
\begin{eqnarray}\label{effectivemass0}
\mu_e=m-\frac{\kappa^2 h}{\omega},~~~\vec{q}=\frac{\vec{k}}{\mu_e}
\end{eqnarray}
 $\mu_e$ being an effective mass.
 It follows that the magnetic fields $B$ and $B^*$ are such that
\begin{eqnarray}\label{magneticfields}
e^*B^*=-\frac{1}{m\omega_0}~~,~~eB=(m-\mu_e)\omega
\end{eqnarray}
The effective mass is then given in function of the magnetic field
by
\begin{eqnarray*}
\mu_e=m-\frac{eB}{\omega}
\end{eqnarray*}
The Hamilton's equations are then
\begin{eqnarray*}\label{hamiltonequationtwofields}
\frac{d\pi_i}{dt}=-\frac{\partial H}{\partial q^i}-
(m-\mu_e)\omega\epsilon_{ik}\frac{\partial H}{\partial
p_k}~~,~~\frac{dx^i}{dt}=\frac{\partial H}{\partial
p_i}+\frac{\epsilon^{ik}}{2m\omega_0}\frac{\partial H}{\partial q^k}
\end{eqnarray*}
The inverse of $\Omega$ is
\begin{eqnarray*}\label{kirillovnewtonhooke2a}
\Omega^{-1}=\left(
\begin{array}{cccc}
0& -
\frac{\omega}{\mu_e}&-\frac{1}{\mu_e}&0\\\frac{\omega}{\mu_e
}&0&0&-\frac{1}{\mu_e}\\\frac{1}{\mu_e}&0&0&\frac{1}{\mu_e\omega_0}\\0&\frac{1}{\mu_e}&-\frac{1}{\mu_e \omega_0}&0
\end{array}
\right)
\end{eqnarray*}
where we have used the wave-particle duality and (\ref{effectivemass0}).
Finally the orbit is equipped with the ( modified) symplectic form
\begin{eqnarray}
\sigma=dp_i\wedge
dq^i+\frac{1}{\mu_e\omega_0}\epsilon^{ij}dp_i\wedge dp_j-
\mu_e\omega \epsilon_{ij} dq^i\wedge dq^j
\end{eqnarray}

We observe that with the planar anisotropic Static group, the phase space obtained is completely noncommutative.
In fact, with the coadjoint orbit
method applied to the centrally extended anisotropic Static Lie algebra, the phase space obtained is such that positions as well as momenta are
noncommutative due to the noncommutativity of both the generators of the
pure Static transformations and the generators of space transformations.  This noncommutativity is measured by two naturally introduced
magnetic fields expressed by relations (\ref{magneticfields}).
The same results have been obtained in the planar anisotropic oscillating Newton-Hooke group case \cite{ancilla}.
\subsubsection{Carroll noncommutative phase spaces}
The planar anisotropic Carroll group is the Carroll group $C(2)$ without the rotations parameters \cite{3derome}.
Its Lie algebra has the only nontrivial Lie bracket given by (\ref{carrolbracket})
 where the left invariant vector fields are given by :
$$\vec{K}=\frac{\partial}{\partial \vec{v}},~~\vec{P}=\frac{\vec{v}}{c^2}\frac{\partial}{\partial t}+\frac{\partial}{\partial \vec{x}},~
H=\frac{\partial}{\partial t}.$$
By using standard methods, we have obtained that the central extension of the planar anisotropic Carroll algebra
is given by (\ref{carrolbrackets}).\\

Let
$hS^*+k_iK^{*i}+p_iP^{*i}+EH^*$ be the general element of
the dual of the centrally extended planar Carroll Lie algebra. Then $E$ and $h$ are
trivial invariants under the coadjoint action. \\
 The restriction of the Kirillov form on the orbit in the basis ( $K_1,K_2,P_1,P_2,H$) is in this case
\begin{eqnarray*}\label{kirillovcarroll}
\Omega=\left(
\begin{array}{cccc}
0&\frac{h}{c^2}&\frac{E}{c^2}&0\\-\frac{h}{c^2}&0&0&\frac{E}{c^2}\\-\frac{E}{c^2}&0&0&\kappa^2 h\\0&-\frac{E}{c^2}&-\kappa^2 h&0
\end{array}
\right)
\end{eqnarray*}
\\
By using relations $h\omega_0=mc^2=E$, $c=\frac{\omega}{\kappa}$ and (\ref{effectivemass0}),
we obtain similar results as in the Static group case.
Thus, by applying coadjoint orbit
method to the extended planar anisotropic Carroll group, the structure of the phase space obtained is the
 same as in the Static group case: positions as well
as momenta do not
noncommute due to the noncommutativity of both the generators of the pure Carroll transformations and the generators of space transformations.
\subsection{ Noncommutative phase spaces on absolute time groups}
For the planar absolute time groups, the noncommutative phase spaces are obtained by
working with the noncentral extensions.
As it has been already said, noncommutative phase spaces on noncentrally extended planar Galilei and Para-Galilei Lie algebras have
been constructed, cf \cite{ancilla1}.
In addition, we have found that noncentral extensions of Newton-Hooke Lie algebras lead to nonvanishing
commutator of two bosts and two momenta as
their corresponding anisotropic Lie algebras \cite{ancilla}.  This means that
totally noncommutative phase spaces can be realized with
 the Newton-Hooke groups in both cases.\\However, noncommutative phase spaces obtained with noncentral extensions of Newton-Hooke groups
are geometrically more general than those obtained with central extensions of the same groups because their symplectic two-forms contain
additional terms.
In this section we construct the noncommutative
phase space on noncentrally extended Static Lie algebra, the only case which have not been done in the
set of absolute time groups.

\subsubsection{Static noncommutative phase space in the absolute time case}
 The noncentrally extended Static Lie algebra of the planar Static group satisfies the nontrivial Lie brackets (\ref{anisotropicextg}).\\
Let
 $$g=exp(v^iK_i+x^iP_i+tH)exp(\theta J)$$
be the general element of the Static group
and $$\hat{g}=exp(\xi M+bS+\varphi M^{\prime}+a\Lambda)exp(\eta^iF_i+{\cal}{l}^i\Pi_i)g$$
be the general element of the connected Lie group associated to the noncentrally extended Static Lie algebra ${\cal{G}}$.
By using the Baker-Hausdorf formulae \cite{hall} and by identifying $\hat{g}$ with $(\beta,\vec{\nu}, g)$
 where $\beta=(\xi,\varphi,b,a)$, $\vec{\nu}=(\vec{\eta},\vec{l})$,
 we obtain that the multiplication law of the corresponding extended Lie group
is
\begin{eqnarray*}
 (\beta,\vec{\nu},g)(\beta^{~\prime},\vec{\nu}^{~\prime},g^{\prime})=(\beta+\beta^{~\prime}+c(g,g^{\prime}),R(\theta)\vec{\nu}^{~\prime}+
\vec{\nu}+\vec{c}~(g,g^{\prime}),gg^{\prime})
\end{eqnarray*}
with
\begin{eqnarray*}\label{aristotelaw}
gg^{\prime}=(\theta,\vec{v},\vec{x},t)(\theta~^{\prime},\vec{v}~^{\prime},\vec{x}~^{\prime},t~^{\prime})=(\theta+\theta~^{\prime},
R(\theta)\vec{v}~^{\prime}+\vec{v},R(\theta)\vec{x}~^{\prime}+\vec{x},t+t~^{\prime})
\end{eqnarray*}
 and
where
\begin{eqnarray*}
c(g,g^{\prime})=(-\frac{1}{2}R(-\theta)\vec{v}.\vec{x}^{~\prime}+\frac{1}{2}R(-\theta)\vec{x}.\vec{v}~^{\prime},
-\frac{1}{2}R(-\theta)\vec{v}.\vec{l}^{~\prime}+\frac{1}{2}R(-\theta)\vec{l}.\vec{v}~^{\prime}\\-\frac{1}{2}\vec{l}.\vec{v}
-\frac{1}{2}R(-\theta)\vec{l}^{~\prime}.R(-\theta)\vec{v}^{~\prime},\\
-\frac{1}{2}R(-\theta)\vec{v}.\vec{\eta}^{~\prime}+\frac{1}{2}R(-\theta)\vec{\eta}.\vec{v}~^{\prime}
-\frac{1}{2}R(-\theta)\vec{x}.\vec{l}^{~\prime}+\frac{1}{2}R(-\theta)\vec{l}.\vec{x}~^{\prime}\\-\frac{1}{2}\vec{\eta}.\vec{v}-\frac{1}{2}
\vec{l}.\vec{x}-\frac{1}{2}R(-\theta)\vec{l}^{~\prime}.R(-\theta)\vec{x}^{~\prime}-\frac{1}{2}R(-\theta)\vec{\eta}^{~\prime}.R(-\theta)
\vec{v}^{~\prime},\\
-\frac{1}{2}R(-\theta)\vec{x}.\vec{\eta}^{~\prime}+\frac{1}{2}R(-\theta)\vec{\eta}.\vec{x}~^{\prime}-\frac{1}{2}\vec{\eta}.\vec{x}
-\frac{1}{2}R(-\theta)\eta^{~\prime}.R(-\theta)\vec{x}^{~\prime})\\
\end{eqnarray*}
while
\begin{eqnarray*}
\vec{c}~(g,g^{\prime})=(\frac{1}{2}[\vec{x}t^{\prime}-tR(\theta)\vec{x}~^{\prime}],\frac{1}{2}[\vec{v}t^{\prime}-tR(\theta)\vec{v}~^{\prime}])
\end{eqnarray*}
 $\theta$ being an angle of rotations, $\vec{v}$  a boost
vector, $\vec{x}$ a space translation vector and $t$ being a time
translation parameter.\\

Let $jJ^*+mM^*+\beta S^*+\mu M^{\prime*}+\kappa\Lambda^*+f_i F^{*i}+I_i\Pi^{*i}+k_iK^{*i}+p_iP^{*i}+EH^*$, ($ i=1, 2$)
be the general element of the dual of the noncentrally extended Static Lie algebra where
the observables are an angular momentum $j$, a static momentum $\vec{k}$, an energy $E$, a force $\vec{f}$, two linear
momenta $\vec{p}$ and $\vec{I}$, two
masses $m$ and $\mu$, a frequency $\omega=\frac{\beta}{\mu}$ and a Hooke's constant $\kappa$.\\

Then  $m,\mu, \beta$ and $ \kappa$ are trivial invariants under coadjoint action of the Static group in two-dimensional space.
The nontrivial invariants $s$ and $U$, solutions of (\ref{kirillovsystem}), are explicitly given by:
% and
%  coadjoint orbit denoted by ${\cal}{O}_{(m,\mu,\beta,\kappa,s,U)}$ is then characterized by four
% trivial invariantsand two given by:
\begin{eqnarray*}
s=j-(\vec{k}-\frac{\beta}{\kappa}\vec{p}~) \times\vec{u}+(\vec{p}-\frac{\beta}{\mu}\vec{k}~) \times\vec{q}\\
 U=E- \frac{\mu_e\vec{u}^{~2}}{2} -\frac{\kappa_e\vec{q}^{~2}}{2}-\frac{\beta\mu_e}{\mu}\vec{q}.\vec{u}-\nu~h
\end{eqnarray*}
where $\nu$ is a frequency and where we have used the relations
\begin{eqnarray*}
 \vec{I}=\mu_e\vec{u},~\vec{f}=-\kappa_e\vec{q},~h=j-s
\end{eqnarray*}
defining the velocity vector $\vec{u}$, the position vector $\vec{q}$ and the action $h$
where
\begin{eqnarray*}\label{effectivehook}
\kappa_e=\kappa-\frac{\beta^2}{\mu},~
\mu_e=\mu-\frac{\beta^2}{\kappa}
\end{eqnarray*}
are the effective Hooke's constant and the effective mass respectively.\\

The inverse $\Omega^{-1}$ of $\Omega$ ( i.e the restriction of the Kirillov form in the basis ($J,P_1,P_2,K_1,K_2,F_1,F_2,\Pi_1,\Pi_2,
H,M,M^{\prime},S,\Lambda$) to the orbit) is
%  is :
% \begin{eqnarray*}\label{kirillovnoncstatic}
% \Omega=\left(
% \begin{array}{cccccccc}
% 0&0&-m&0&\kappa&0&\beta&0\\
% 0&0&0&-m&0&\kappa&0&\beta\\
% m&0&0&0&\beta&0&\mu&0\\
% 0&m&0&0&0&\beta&0&\mu\\
% -\kappa&0&-\beta&0&0&0&0&0\\
% 0&-\kappa&0&-\beta&0&0&0&0\\
% -\beta&0&-\mu&0&0&0&0&0\\
% 0&-\beta&0&-\mu&0&0&0&0\\
% \end{array}
% \right)
% \end{eqnarray*}

\begin{eqnarray*}
\Omega^{-1}=\frac{1}{\beta^2-\mu\kappa}\left (\begin{array}{cccccccc}
0&0&0&0&\mu&0&-\beta&0\\
0&0&0&0&0&\mu&0&-\beta\\
0&0&0&0&-\beta&0&\kappa&0\\
0&0&0&0&0&-\beta&0&\kappa\\
-\mu&0&\beta&0&0&0&m&0\\
0&-\mu&0&\beta&0&0&0&m\\
\beta&0&-\kappa&0&-m&0&0&0\\
0&\beta&0&-\kappa&0&-m&0&0\\
\end{array}
\right)
\end{eqnarray*}
We then verify that the symplectic form on the orbit  ${\cal}{O}_{(m,\mu,\beta,\kappa,s,U)}$
 is
\begin{eqnarray}\label{symplecticformncstatic2}
\sigma= d{\vec{p}}\wedge d{\vec{q}}+ ~d{\vec{k}}\wedge d{\vec{u}}+\frac{\beta}{\kappa}~ d{\vec{p}}\wedge {d \vec{u}}
-\frac{\beta}{\mu} ~d{\vec{k}}\wedge d{\vec{q}}
\end{eqnarray}
The Poisson brackets of two functions $H$ and $f$ on the orbit \\corresponding to the symplectic form
 (\ref{symplecticformncstatic2}) is
\begin{eqnarray*}
\{H,f\}=\frac{\partial H}{\partial p_i}\frac{\partial f}{\partial q^i}-\frac{\partial H}{\partial q^i}\frac{\partial f}{\partial p_i}+
\frac{\partial H}{\partial k_i}\frac{\partial f}{\partial u^i}-\frac{\partial H}{\partial u^i}\frac{\partial f}{\partial k_i}\\-
\frac{\beta}{\kappa}{\epsilon_{i}}^j\frac{\partial H}{\partial p_i}\frac{\partial f}{\partial u^j}+
\frac{\beta}{\mu}{\epsilon^{i}}_j\frac{\partial H}{\partial q^i}\frac{\partial f}{\partial k_j}
\end{eqnarray*}
and the nontrivial Poisson brackets within the coordinates are
\begin{eqnarray*}\label{possonstaticabs}
\{p_j,q^i\}=\delta^i_j~~,~\{k_j,u^i\}=\delta^i_j~,~\{p_j,u^i\}=\frac{\beta}{\kappa}{\epsilon_{i}}^j~,\\~\{q^i,k_j\}=\frac{\beta}{\mu}
{\epsilon^{i}}_j,
~\{p_i,k_j\}=0,~\{q^i,u^j\}=0
\end{eqnarray*}
This means that the linear momentum $\vec{p}$ is canonically conjugate to the position $\vec{q}$ as well as the static momentum $\vec{k}$
is canonically conjugate to the the velocity $\vec{u}$.  Moreover the linear momentum does not commute with the velocity as well as the static
 momentum does not commute with the position.
Note that in this case, $(\vec{q},\vec{u})$ is an element of the tangent space (evolution space) while $(\vec{p},\vec{k})$ is its dual.
\subsubsection{Symplectic realizations and equations of motion}

 Let the symplectic realizations of the noncentrally extended Static group on its coadjoint orbit be given by
$(\vec{p}^{~\prime},\vec{k}^{~\prime},\vec{q}^{~\prime},\vec{u}^{~\prime})=D_{(\vec{\eta},{\cal}{\vec{l}},\theta,\vec{v},\vec{x},t)}
(\vec{p},\vec{k},\vec{q},\vec{u})$ .
 By using coadjoint action, we verify that
\begin{eqnarray*}
\vec{q}^{~\prime}=R(\theta)\vec{q}+\vec{v}\tau+\frac{\kappa}{\kappa_e}\vec{x}\\
\vec{u}^{~\prime}=R(\theta)\vec{u}-\nu_e\vec{x}-\frac{\mu}{\mu_e}\vec{v}\\
\vec{p}^{~\prime}=R(\theta)\vec{p} -t\kappa_e [R(\theta)\vec{q}-\vec{v}\tau -\frac{\kappa}{\kappa_e} \vec{x}~]-m\vec{v}+\kappa \vec{\eta}
+\beta \vec{l}\\
\vec{k}^{~\prime}=R(\theta)\vec{k}+t\mu_e [R(\theta)\vec{u}-\nu_e\vec{x}- \frac{\mu}{\mu_e} \vec{v}~]+m \vec{x}+\mu \vec{l}+\beta \vec{\eta}
\end{eqnarray*}
$\tau=\frac{\beta}{\kappa_e},~\nu_e=\frac{\beta}{\mu_e}$ being a duration and a frequency respectively.\\

 The evolution with respect to the time  $t$ is
\begin{eqnarray*}\label{configurationevolution2st}
\vec{q}~(t)=q~,~\vec{u}~(t)=\vec{u}
\end{eqnarray*}
and
\begin{eqnarray*}\label{momentaevolution2st}
\vec{p}~(t)=\vec{p}-t\kappa_e\vec{q}\\
\vec{k}~(t)=\vec{k}+t\mu_e\vec{u}
\end{eqnarray*}

The equations of motion are then given by
\begin{eqnarray}\label{positionimpulse}
\frac{d}{dt}\left(
\begin{array}{c}
\vec{p}(t)\\ \vec{k}(t)
\end{array}
\right)=\left(
\begin{array}{c}
-\kappa_e\vec{q}\\\mu_e\vec{u}
\end{array}
\right)~,~
\frac{d}{dt}\left(
\begin{array}{c}
\vec{q}(t)\\ \vec{u}(t)
\end{array}
\right)=\left(
\begin{array}{c}
0\\0
\end{array}
\right)
\end{eqnarray}
\\
With the planar noncentrally extended Static group, we have obtained a completely noncommutative phase space equipped with
 modified symplectic structure defined by
 relation (\ref{symplecticformncstatic2}) and dynamics given by equations (\ref{positionimpulse}).\\

We can summarize our findings.  From the group theoretical discussion above, we see that the coadjoint orbit method applied to
the two-parameter central extensions of anisotropic Lie groups and to
the noncentral extensions of
absolute time Lie groups gives rise to three kinds of noncommutative phase spaces :
\begin{itemize}
 \item phase spaces whose only positions do not commute, i.e : generators of pure kinematical group transformations are noncommutative in
extended groups. This arrives with
the Galilei algebra $ G$ by considering both the anisotropic and the absolute time cases.
\item phase spaces whose only momenta are noncommutative due to the noncommutativity of generators of space translations in
extended Lie groups.
This is the case of the Para-Galilei group $ G^{\prime}$ by considering both the anisotropic and the absolute time cases.
\item phase spaces which are completely noncommutative: that means that both generators of pure kinematical group transformations and generators
of space translations do not commute. These are obtained with the Newton-Hooke Lie groups $ NH_{\pm}$,
the Static Lie group and the Carroll Lie group.\\
\end{itemize}
% Thus,, we see that the noncommuting spatial coordinates or positions (only) arise from the Galilei group,
%  the noncommuting momenta (only) arise from the Para-Galilei group while the noncommuting positions and momenta (both) arise from
% Newton-Hooke, Static and
% Carroll groups
% in the anisotropic case (with the two-parameter centrally extended group) and in the absolute time case (with the noncentrally extended group).\\

%  in the anisotropicand its
% noncentrally extended group
% it is the Galilei group (by considering its two-fold centrally ) which is responsible to the noncommutativity
% \end{itemize}

{\bf {Aknowledgements}}
The authors are grateful to Professor Peter Horvathy for useful remarks which allowed us to improve the paper.

 \end{document}